\documentclass[preprint]{aastex}
\usepackage{amssymb,bm,graphicx,rotating}

\newcommand{\beq}{\begin{equation}}
\newcommand{\eeq}{\end{equation}}
\newcommand{\dee}[1]{\,{\rm d}{#1}}

\shorttitle{Shock generated vorticity and the IMF}
\shortauthors{N. Kevlahan}

\begin{document}
\title{\vspace*{-2.5cm}Shock generated vorticity in the interstellar medium \\ and 
the origin of the stellar initial mass function} 
\author{N. Kevlahan}
\email{kevlahan@mcmaster.ca}
\affil{Department of Mathematics \& Statistics, McMaster University,
Hamilton, ON L8S 4K1, Canada}
\affil{Origins Institute, McMaster University,
Hamilton, ON L8S 4M1, Canada}

\and

\author{Ralph E. Pudritz}  
\affil{Origins Institute, McMaster University,
Hamilton, ON L8S 4M1, Canada}

\begin{abstract}
Observations of the interstellar medium (ISM) and molecular clouds
suggest these astrophysical flows are strongly turbulent.  The main
observational evidence for turbulence is the power-law energy spectrum
for velocity fluctuations, $E(k)\propto k^\alpha$, with $\alpha\in
[-1.5,-2.6]$.  The Kolmogorov scaling exponent, $\alpha=-5/3$, is
typical.  At the same time, the observed probability distribution
function (PDF) of gas densities in both the ISM as well as in
molecular clouds is a log-normal distribution, which is similar to the
initial mass function (IMF) that describes the distribution of stellar
masses.

In this paper we examine the density and velocity structure of
interstellar gas traversed by curved shock waves in the kinematic
limit.  We demonstrate mathematically that just a few passages of
curved shock waves generically produces a log-normal density PDF.
This explains the ubiquity of the log-normal PDF in many different
numerical simulations.  We also show that subsequent interaction with
a spherical blast wave generates a power-law density distribution at
high densities, qualitatively similar to the Salpeter power-law for
the IMF.  Finally, we show that a focused shock produces a {\em
downstream\/} flow with energy spectrum exponent $\alpha=-2$.
Subsequent shock passages reduce this slope, achieving $\alpha\approx
-5/3$ after a few passages.  We argue that subsequent dissipation of
energy piled up at the small scales will act to maintain the spectrum
very near to the Kolomogorov value despite the action of further
shocks that would tend to reduce it.  These results suggest that
fully-developed turbulence may {\em not\/} be required to explain the
observed energy spectrum and density PDF.

On the basis of these mathematical results, we argue that the
self-similar spherical blast wave arising from expanding HII regions
or stellar winds from massive stars may ultimately be responsible for
creating the high mass, power-law, Salpeter-like tail on an otherwise
a log-normal density PDF for gas in star forming regions.  The IMF
arises from the gravitational collapse of sufficiently overdense
regions within this PDF.  Thus, the composite nature of the IMF --- a
log-normal plus power-law distribution --- is shown to be a natural
consequence of shock interaction and feedback from the most massive
stars that form in most regions of star formation in the galaxy.

\end{abstract}


\section{Introduction\label{sec:intro}}

The distribution of stellar masses, known as the initial mass function
(IMF), is of paramount importance to many fields of astrophysics.  The
form of the IMF plays a central role in subjects as diverse as
galactic evolution and the formation and evolution of exoplanetary
systems.  Although its origin has been the focus of many analytical
models for several decades, only recently have numerical simulations
become available that can include many of the important physical
processes involved \citep[see the reviews by][]{McKee/Ostriker:2007,
Bonnell/etal:2007}.  The form of the IMF is variously described as a
log-normal distribution at low stellar masses with a power-law tail at
masses exceeding a solar mass~\citep[e.g.][]{Chabrier:2003}, or as a
multiple power-law \citep[e.g.][]{Kroupa:2002}.  The Salpeter
power-law index of -1.35 for the high mass power-law tail appears to
be universal.  These properties of the IMF must ultimately reflect
robust properties of the dense substructure within molecular clouds in
the interstellar medium (ISM), in which stars are born.

It is not surprising therefore that one of the central problems in
star formation is to characterize the mass distribution of the dense
star-forming regions within clouds.  Millimetre and submillimetre wave
observations show that these clouds are highly inhomogeneous and are
dominated by systems of filaments, punctuated by smaller denser
regions in which clusters of stars form.  Individual stars form in
dense regions ($n\ge 10^4 \mathrm{cm}^{-3}$ ), whose mass distribution
is known as the core mass function (CMF).  The gas velocities in these
clouds are observed to be supersonic and chaotic.  Numerical studies
reveal that ``turbulent" supersonic gas motions can reproduce many
aspects of this structure.  A number of observational surveys have
shown that the CMF can be modelled by a log-normal distribution in
many instances
\citep[e.g.][]{Goodman/etal:2008}.  Other studies suggest that the high mass
tail of this distribution is closer to a power-law, whose index is
nearly identical to the Salpeter value~\citep[e.g.][]{Motte/etal:1998,
Johnston/etal:2000}.  However, the debate continues as to the exact
form of the mass distribution of these dense gaseous structures
\citep[e.g.][]{Goodman/etal:2008}.

This subject has made major advances largely due to the advent of
numerical simulations.  These show that the filamentary nature and the
mass spectra of structure in clouds naturally result from the action
of supersonic turbulence within them \citep[e.g., the review of][]{
MacLow/Klessen:2004}. The precise origin of the turbulence is still
somewhat unclear.  Molecular clouds are themselves embedded in a
multi-component ISM and could be formed by several processes,
including cloud collisions, spiral shock waves, and a combination of
gravitational and magnetic instabilities.  They are also shocked by
the effects of feedback from star formation itself including expanding
HII regions, supernovae, and powerful winds from massive stars.
Simulations have started to make substantial progress in following
most of these processes \citep[e.g.][]{Wada/Norman:2001,
Tasker/Bryan:2006}.  The presence of supersonic turbulence within all
molecular clouds has been interpreted as evidence that they are short
lived structures that are dissipated in one or two shock crossing
times: star formation occurs in the few transient turbulent structures
that have sufficiently high density to collapse
\citep[e.g.][]{Elmegreen:2002, Hartmann/etal:2001}.  
Indeed, observational studies of molecular clouds
in the nearby LMC galaxy indicate that the time scale to form star clusters
is rapid (of order a Myr) and that molecular clouds are rapidly dissipated
in a few Myr as a consequence of their formation \citep[][]{Fukui/etal:1999}.  
Turbulence within such clouds may result from processes on
larger scales ($>100\mathrm{pc}$) that tap into energy released by galactic
shear, gravitational instability or large-scale expanding HII regions.

In spite of the wide range of conditions in the interstellar medium and
the details of the modelled process, time-dependent numerical simulations show 
that the density structure can be well modelled
by a log-normal PDF over several orders of magnitude: from the diffuse
atomic gas to molecular clouds \citep[eg.][]{Wada/Norman:2007, Tasker/Bryan:2008}.
At later times in their simulations of the ISM undergoing feedback from the effects
of massive star formation \citet{Tasker/Bryan:2008} found that a power-law
fit might also be possible.  Numerical
simulations of density fluctuations in purely isothermal supersonic
turbulence have a log-normal PDF, and this is often taken as evidence
for turbulence in the ISM.  However, \citet{Padoan/Nordlund:2002} used
the central limit theorem to show that a flow with a power-law energy
spectrum will necessarily have a log-normal PDF of density. Thus, a
log-normal PDF of density does not provide any more evidence for
turbulence than a power-law energy spectrum.  

How does the distribution of stellar masses arise from this density structure
that characterizes the ISM?  Stars are formed in  fluctuations 
whose density exceeds a threshold for gravitational collapse. Recent theoretical work
has emphasized that the shape of the IMF may be a combination of a
power-law at large mass scales, that transitions to a log-normal form
at lower masses.  The peak of this distribution is a mass
characteristic of gravitational collapse
\citep[e.g.][]{Hennebelle/Chabrier:2008,Padoan/Nordlund:2002}.
The power-law tail and near Salpeter index for the mass function in
molecular clouds has been modelled as arising from shocks generated in
nearly Kolomogorov turbulence \citep[e.g.][]{Padoan/Nordlund:2002},
where the power-law index depends on the spectral index of the
turbulent flow, and the half-width of the log-normal distribution
depends on the turbulence Mach number.  A recent analytic approach
argues that the log-normal distribution for the IMF at low masses
arises from that part of the gas density distribution that is
supported by thermal pressure, whereas the Salpeter tail arises for
higher mass cores that are supported by turbulent pressure
\citep[]{Hennebelle/Chabrier:2008}.  An alternative explanation for a
joint log-normal plus power-law distribution initial distribution is
that an initial log-normal distribution can develop a power law tail
if cores accrete over a distribution of time scales
\citep[]{Basu/Jones:2004}.  Finally, \citet{Elmegreen:2002} noted that
if one assumes that all gas above a density threshold in a log-normal
distribution can form stars, then it is possible to recover the
well-known Schmidt law that governs the global star formation rate in
galaxies \citep[see also][]{Wada/Norman:2007}.  .

In this paper, we examine the mathematical properties of shock-driven
gas motions and propose a new approach to explain the nature of gas
motions and density structure in the ISM and molecular clouds.  We use
analytical theory to examine both the density distribution expected in
the gas due to the interaction of curved shocks, as well as the nature
of the velocity fluctuations downstream of a shock.  We show that the
passage of just a few shock waves can very quickly establish
log-normal density distributions.  The passage of a spherical shock
(i.e. blast wave) adds a power-law tail at large mass densities.  We
demonstrate that in general a log-normal distribution with a power-law
at large densities is expected in media that are occasionally
traversed by such large scale spherical shocks. These spherical shocks
have been observed in the expanding HII regions that are a consequence
of massive star formation.  Thus, although gravity may be important
for the large scale dynamics of molecular clouds
\citep[][]{Goodman/etal:2008}, the large scale shock waves of varying
symmetry play an essential role in shaping the mass function of the
cores.

The motions induced in the wake of curved shocks are vortical in
nature.  Power is distributed to vortical motions across a wide range
of scales without a cascade process that is essential for Kolomogorov
turbulence.  Although tempting, we will argue that it is problematic
to interpret the observations of Kolmogorov-like scaling in terms of
hydrodynamic (or MHD) turbulence.  

The paper is organized as follows.  In \S\ref{sec:shock} we contrast
the properties of shock-driven and Kolmogorov turbulence. Then in
\S\ref{sec:vort} we review
\citet{Kevlahan:1997}'s theory for vorticity generation by shocks
propagating in nonuniform flows. Although it is commonly thought that
straight shocks, weak shocks, and spherical shocks do not generate
vorticity~\citet{Kevlahan:1997} demonstrated that this is not true for
shocks in inhomogeneous flows.  In addition, we highlight the fact
that curved shocks eventually focus.  This focusing produces a pair of
shock-shocks~\citep{Whitham:1974} which generate vortex sheets
downstream of the shock. This effect appears not to have been
considered before in astrophysical flows.  In \S\ref{sec:density} we
show how multiple shock interactions could generate the observed
log-normal and power-law distributions of mass density, and in
\S\ref{sec:spec} we show that the observed velocity energy spectra
could be produced by the quasi-singular vorticity generated downstream
of focused shocks and blast waves.  Finally, we interpret the results in 
terms of an astrophysical model for the role of shock waves in generating
density structure in the ISM and molecular clouds, and its connection to the IMF
and close with our conclusions for shocks and star formation in the ISM
\S\ref{sec:concl}.

\section{Shock-driven flow and Kolmogorov turbulence\label{sec:shock}}
Observations of density and velocity fluctuations have suggested that
many astrophysical flows are strongly turbulent.  This phenomenon is
widespread and includes a diverse set of systems --- including H~I
emission in the interstellar medium (ISM), interstellar scintillations
(ISS), 100$\mu$ IRAS emission in the interstellar medium
(ISM)~\citep{Elmegreen/Scalo:2004}, velocity and density structure in
molecular clouds, H~I emission in the large Magellanic cloud
(LMC)~\citep{Elmegreen/etal:2001}, fluctuations in the solar
wind~\citep{Horbury:1999,Nicol/etal:2008}, and hot H${_2}0$ emission
in accretion disks~\citep{Carr/etal:2004}.  The evidence suggesting
that these fluctuations are in fact turbulent is principally their
turbulent-like velocity dispersions, or, equivalently, energy spectra
and second-order structure functions, as well as their spatial density
structure.  Sometimes the scaling of higher-order structure functions
is also taken as evidence of turbulence
\citep[e.g.][]{Padoan/etal:2003}.  However, these results are not
conclusive since there is still no rigorous theory for the scaling of
high-order structure functions \citep[only models, such as the one proposed
by][]{She/Leveque:1994}, and there is not enough data to obtain
properly converged statistics for structure functions of order greater than
about 6 or 7 without using the extended self-similarity (ESS) correction.

One of the central ideas in Kolmogorov's classical theory of
turbulence is that energy is injected at large scales and cascades
without loss through intermediates scales, until it reaches the
smallest scales where it is finally lost to molecular dissipation
\citep[see][for a detailed review]{Elmegreen/Scalo:2004}.  Fully 
developed incompressible turbulence is characterized by a power-law
energy spectrum $E(k)\propto k^\alpha$ where $\alpha=-5/3$ for incompressible
homogeneous isotropic three-dimensional turbulence and
$\alpha\in[-3,-4]$ for the enstrophy cascade in two-dimensional
turbulence.  Astrophysical fluctuations have observed power-law
spectra with $\alpha\in[-1.5,-2.6]$ \citep{Elmegreen/Scalo:2004},
suggestive of turbulence.  Radio propagation observations of the
diffuse ISM have found that density fluctuations obey the ``Big power
law in the sky'': Kolmogorov-like scaling of the energy spectrum
extends over 11 orders of magnitude, from $10^7$cm to $10^{18}$ cm
\citep{Spangler:1999}!   However,
\citet{Lazarian/Pogosyan:2000} comment that ``the explanation of the spectrum as due
to a Kolmogorov-type cascade faces substantial difficulties.''
Indeed, they emphasize that ``The existence of the Big Power Law
\citep[see][]{Armstrong/etal:1995, Spangler:1999} is one of the great
astrophysical mysteries.''

Measurements of the velocity dispersion of gas within molecular clouds
are made as a function of the size of the region in which the
dispersion is measured.  The data show that there is a power-law
relation between the linewidth and size of a region such that $\Delta
v \propto R^{\beta}$ \citep{Ballesteros-Paredes/etal:2007}.  This
scaling is a direct consequence of the energy spectrum since $\beta =
-(\alpha +1)/2 $.  \cite{Larson:1981} first deduced that $\beta
\simeq 0.38$, which is close to Kolmogorov turbulence $ (\beta = 1.3,
\alpha = -5/3)$, and was the first to suggest that turbulence must
play a very important role in the gas, and as a consequence, in the
process of star formation that occurs in such clouds.  Recent surveys
for whole giant molecular clouds (GMCs) find $\beta \simeq 0.5-0.6$
\citep[e.g.][]{Heyer/Brunt:2004}, while studies of clouds with regions of
low surface brightness find $\beta \simeq 0.4$
\citep{Falgarone/etal:1992}.  Compressible gas motions which
characterize the ISM are damped very quickly in shocks --- typically in
one crossing time of the driving scale.

Turbulence can also be characterized by the scaling of $\zeta_p$, the
exponent of the $p-$th order structure function,
\begin{equation}
\langle |u(x+r) - u(x)|^p \rangle \propto r^{\zeta_p}.
\end{equation}
Kolmogorov's theory~\citep{Frisch:1995} predicts that $\zeta_p$ is a
linear function of $p$ (with $\zeta_p=p/3$). However, experiments show
that $\zeta_p$ is in fact a concave function of $p$, increasing more
slowly than linear with order $p$.  Some attempts have been made to
measure structure function exponents for astrophysical
fluctuations~\citep[e.g.][]{Nicol/etal:2008}, although lack of data
restricts the analysis to relatively low order ($p\le 4$ or $6$) and
quantitative comparison with turbulent flows is difficult since there
is no accepted theory for how $\zeta_p$ should scale.  

The equation of spectral energy balance for approximately incompressible homogeneous decaying turbulence is
\[
\partial_t E(k) = -2\nu k^2 E(k) + T(k),
\]
where $k=|\bm{k}|$ is the magnitude of the wavenumber, the first term
on the right hand side is viscous dissipation of energy (active only
at small scales) and $T(k)$ measures the rate of energy transfer from
all other wavenumbers $k'$ to $k$ due to nonlinear interactions,
i.e. it quantifies the energy cascade.  Thus, in order to demonstrate
conclusively the existence of an energy cascade one must estimate the
energy transfer function,
\[
T(k)=\int_{k=|\bm{k}|} \hat{\bm{u}}^*(\bm{k})\cdot \bm{P}(\bm{k})(\widehat{\bm{u}\times\bm{\omega}}(\bm{k}))\;\mathrm{d} S(k),
\]
where $\hat{(\hspace{1ex})}$ denotes the Fourier transform,
$\bm{\omega}$ is the vorticity, $\bm{P}(\bm{k})$ is the
divergence--free projection (for approximately incompressible flow)
and the integral is over spherical shells in Fourier space.  However,
it is impossible to estimate $T(k)$ from the available the
observational data since calculating $\bm{u}\times\bm{\omega}$
requires pointwise measurements of all components of the velocity and
vorticity.

The interpretation of power-law scaling of the energy spectrum in
terms of fully developed Kolmogorov turbulence in the ISM is
problematic for several reasons.

\begin{enumerate}
\item
Kolmogorov scaling is associated with incompressible neutral flow,
whereas the ISM is believed to be strongly compressible and
magnetic. \citet{Sridhar/Goldreich:1994} proposed a theory for
anisotropic incompressible magnetohydrodynamic (MHD) turbulence which
gives an energy spectrum $k_{\perp}^{-5/3}$ in directions
perpendicular to the mean magnetic field.  However, the theory is not
rigorous and it does not apply to the solar wind.  Other weak
turbulence calculations find $k_{\perp}^{-2}$ or $k_{\perp}^{-3/2}$,
and stationary constant flux solutions may have exponents anywhere in
the range from $-1$ to $-3$ depending on the asymmetry of the
forcing~\citep{Galtier/etal:2002}.

\item
The ``Big power law in the sky'' extends over a range of
scales that include many different physical processes, including
scales where the gas dynamics approximations of fluid turbulence are
not valid. How can the same scaling be maintained across scales with
very different physics?

\item
It is not clear where the energy sustaining the ISM turbulence comes
from. Candidates include massive stellar winds, supernovae, expanding
HII regions, galactic rotation via spiral shocks, sonic reflection of
shock waves hitting clouds, cosmic ray streaming, field star motions,
Kelvin-Helmholtz and other fluid instabilities, thermal instabilities,
gravitational instabilities, and galaxy
interactions~\citep{Elmegreen/Scalo:2004}.

\item
As pointed out by \citet{Lazarian/Pogosyan:2000}, the damping rate of
MHD turbulence is much faster than previously thought: about one eddy
turnover time (as for neutral fluids).  This implies very large energy
injection scales and efficient and frequent forcing in order to
sustain the turbulence (since the energy cascade takes about one eddy
turnover time).  It is useful to recall that hydrodynamic turbulence
typically has constant or frequent forcing, e.g. vorticity generation
via the no-slip boundary condition, or the mixing layer instability in
jets.  Vorticity generation occurs on a huge range of scales and does
not require the lengthy process of an energy cascade.
\end{enumerate}

In numerical simulations, on the other hand, turbulence is often
generated via spiral shocks or super novae explosions
\citep[e.g.][]{Joung/etal:2006,Piontek/Ostriker:2005,Wada/etal:2002}.  This
turbulence is necessarily limited to low Reynolds numbers, and the
power-law scaling of the energy spectrum is present over a small range
of scales (about a decade). Supersonic turbulence from these and other
simulations of the ISM have a scaling close to the $-2$ associated
with the shock discontinuity~\citep{Kritsuk/etal:2007,Vazquez/etal:1997}.

Mathematically, the energy spectrum of a field is determined by its
strongest singularity.  If a function $f(x)$ has a discontinuity in
the $p-1$ order derivative (where $p$ is an integer), then the energy
spectrum of $f(x)$ has the form of a power-law
\[
E(k)\sim k^{-2p}.
\]
For example, a field containing shocks (discontinuities in the
velocity field) has $E(k)\sim k^{-2}$.  A singularity `worse' than a
discontinuity would be required to generate a slope shallower than
$-5/3$, for example an {\em accumulation\/} of discontinuities around
a given point, or a fractal \citep[see][]{Hunt/etal:1990}.  Note,
however, that the converse is not true: smooth fields can have a
power-law energy spectrum (e.g. adding together Gaussian functions of
just the right size and amplitude could produce a $k^{-5/3}$
spectrum).

The fact that a power-law scaling is observed over such a wide range
of scales suggests that a singularity may be responsible, such as the
shocks that are ubiquitous in astrophysical flows.
\citet{Kornreich/Scalo:2000} have proposed galactic shocks propagating
through interstellar density fluctuations as a way of forcing (or
``pumping'') supersonic turbulence.  We go one step further:
shock-generated vorticity alone may be enough to explain the observed
energy spectra.  Dynamical turbulence is not required.

As mentioned above, simulations of supersonic turbulence typically
produce a $-2$ spectrum.  This is not surprising since flows with
turbulence Mach number $M_t>0.3$ will spontaneously generate
shocklets~\citep{Kida/Orszag:1990}.  Since the Kolmogorov $-5/3$
spectrum is likely not associated with a singularity, the $-2$ scaling
dominates.  However, as pointed out by \citet{Elmegreen/Scalo:2004},
the observed spectra are often shallow than $-2$ and ``the proposed
shocks themselves have not been observed in real clouds.''  In this
paper we suggest a shock-based explanation of the observed spectra
that addresses both these issues.  First, we show that focused shocks
generate {\em vortex sheets\/}, which means that the shock-driven flow
may have a spectrum $E(k)\sim k^{-2}$ even when the shock is no longer
present. Secondly, we show that multiple passages of curved shocks
produces velocity fluctuations with a spectrum shallower than $-2$.
Focused shocks are not the only shock structures that produce
power-law scaling: we also find that multiple passages of strong
spherical shocks produce similar results.

\cite{Dobbs/Bonnell:2007} have recently proposed a similar shock-based
explanation of the velocity dispersion (i.e. energy spectrum) in
molecular clouds.  They use full smoothed particle hydrodynamics (SPH)
simulations of the gas dynamics equations to show that shocks
propagating through non-uniform fractal gas generate velocity
dispersions close to the observed scaling.  Our study is
complementary: we use the full analytic expression for the vorticity
produced by a curved shock in nonuniform flow to show that multiple
shock passages could produce the observed velocity dispersion (and PDF
of density fluctuations).  We identify baroclinic vorticity generation
as the key term.  In addition, we find that fractal (or even
non-uniform) initial conditions are not necessary.

\section{Vorticity generation by curved and oblique shocks\label{sec:vort}}
In this section we review \cite{Kevlahan:1997}'s theory for the
vorticity jump across a shock in non-uniform flow.  We emphasize two
effects that are usually neglected in discussions of vorticity
generation by shocks: the creation of vortex sheets downstream of
highly curved shock regions (e.g. shock--shocks), and the role played
by non-uniformities in the flow ahead of the shock (which can lead to
significant vorticity generation even by straight shocks).

The general expression for the vorticity jump in the binormal
direction $\mathbf{b}$ across an unsteady three-dimensional shock
moving into a non-uniform flow was derived by~\cite{Hayes:1957},
\beq
\delta {\omega} \, \mathbf{b} = \mathbf{n} \times \left[ -\frac{\partial(\rho
C_r)}{\partial S}\delta(\rho^{-1})
(\rho C_{r})^{-1} (D_{S} \mathbf{U}_{S} + C_{r} D_{S}\mathbf{n})\delta (\rho) \right],
\label{eq:vort1}
\eeq
where $\bm{n}$ is the shock-normal direction, $\bm{s}$ is the
tangential direction, and the binormal direction is given by
$\bm{b}=\bm{s}\times\bm{n}$.  Note that both $\bm{s}$ and $\bm{b}$ are
tangential to the shock surface, and the normal component of the
vorticity is continuous across a shock. $\partial/\partial S$ is the
tangential part of the directional derivative, $C_{r}=C-A$ is the
shock speed relative to the normal component of the flow ahead of the
shock $A$, and $D_{S}$ is the tangential part of the total time
derivative.  A similar expression may be derived for the vorticity
jump in the tangential direction $\mathbf{s}$.  However, the
expression usually taken for the vorticity jump is
\beq
\delta \omega \, \mathbf{b} = -\frac{\mu^{2}}{1+\mu} \mathbf{n} \times 
\left( \mathbf{U}_{S}\cdot\mathbf{K} + \frac{\partial C_{r}}{\partial S} 
\right)\mathbf{s}, \label{eq:vort1a}
\eeq
where $\mu$ is the normalized density jump across the shock (the shock
strength), $\mathbf{K}$ is the curvature of the shock and
$\mathbf{U}_S$ is the velocity tangential to the shock in the
reference frame of the shock.  This expression was derived by Hayes
from (\ref{eq:vort1}) by assuming that the flow ahead of the shock is
{\it uniform\/}.

\cite{Kevlahan:1997} re-derived the vorticity jump equation, taking
full account of terms due to the non-uniform upstream, finding that the
vorticity jump in the binormal direction $\bm{b}$
\beq
\delta\bm{\omega}\cdot\bm{b} = 
\frac{\mu^2}{1+\mu} \frac{\partial C_r}{\partial S} -\frac{\mu}{C_r} \left(\left[\frac{D\bm{u}}{Dt} +
\frac{C_r^2}{1+\mu} \frac{1}{\rho} \nabla\rho
\right]\cdot\bm{s}\right) + \mu\bm{\omega}\cdot\bm{b},
\label{eq:voder6}
\eeq
together with a similar expression for the vorticity jump in the
tangential direction $\bm{s}$.  If the upstream flow is isentropic
then $a_0^2/\rho\nabla\rho\approx-D\bm{u}/Dt$, and if, in addition,
the upstream flow is quasi-steady and we normalize by the stagnation
sound speed $a_0$ then (\ref{eq:voder6}) becomes
\beq
\delta\bm{\omega}\cdot\bm{b} = \frac{\mu^2}{1+\mu} \frac{\partial M_s}{\partial S}
+ \frac{1}{M_s}\left(\frac{\mu}{1+\mu} M_s^2-1\right) \left[\frac{\partial\frac{1}{2}M_t^2}{\partial S}
+ \bm{\omega}\times\bm{u}\cdot\bm{s}\right] + \mu\bm{\omega}\cdot\bm{b},
\label{eq:voder7}
\eeq
where $M_t$ is the turbulent Mach number of the upstream flow.  We
will use (\ref{eq:voder7}) in the remainder of the paper. Equation
(\ref{eq:voder7}) may be simplified further for strong shocks by using
the approximation $\mu\approx 2/(\gamma-1)$.

The density jump is given by 
\beq
\delta\rho \equiv \mu\rho = \frac{M_s^2-1}{1+1/2(\gamma-1)M_s^2}\rho,
\label{eq:density}
\eeq
where $\mu$ is often referred to as the shock strength.

The first term on the right hand side of (\ref{eq:voder7}) represents
vorticity generation due to the variation of the shock speed $M_s$
along the shock; by symmetry it is exactly zero for spherical and
cylindrical shocks.  Because shocks are nonlinear waves (unlike
acoustic waves), $M_s$ is larger in regions of concave curvature and
smaller in regions of convex curvature (with respect to the
propagation direction of the shock).  This difference in shock
strength increases over time and eventually causes curved shocks to
focus at regions of minimum curvature, developing a flat shock disk
bounded by regions of very high curvature (often called
kinks)~\citep{Kevlahan:1996}. In laboratory experiments shock focusing
is obtained by reflecting a straight shock off a curved
surface~\citep[see][for a detailed discussion and pictures of shock
focusing experiments]{Sturtevant/Kulkarny:1976}.  The discontinuous
shock strength at the kinks is called a shock--shock. In the ISM shock
focusing could arise due to reflection off density gradients
(e.g. vertically stratified structure in disks), or due to small
variations in shock curvature in blast waves.

\begin{figure}
\[\includegraphics[width=0.6\textwidth,clip=false]{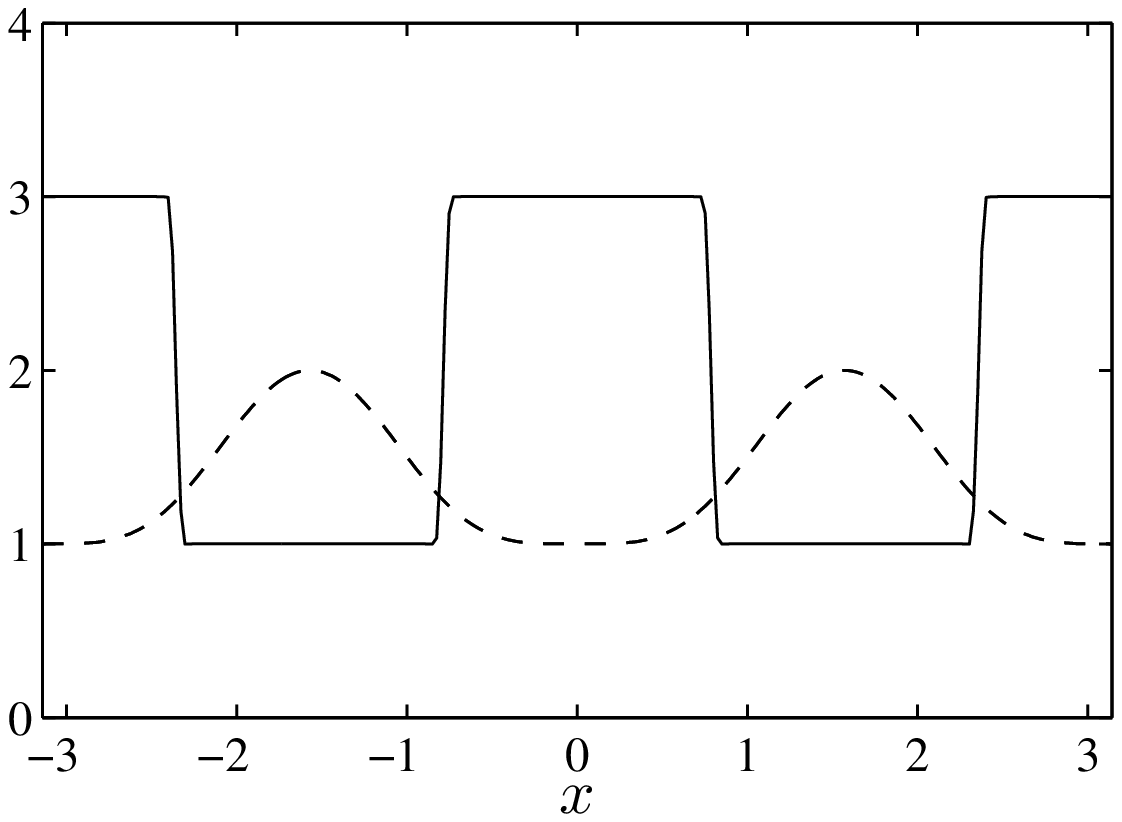}\]
\caption{Focused shock model: --- shock strength, - - - shock
profile.  Note the shock--shocks (discontinuities in shock strength)
which generate vortex sheets in the downstream
flow. \label{fig:focus}}
\end{figure}

As mentioned above, since the first term on the right hand side of
(\ref{eq:voder7}) is approximately singular at the location of a kink,
extremely strong jet-like vortex sheets develop in the flow downstream
of the kinks (see figure~\ref{fig:focus}).  These vortex sheets
themselves have an energy spectrum $E(k)\sim k^{-2}$, and generate
turbulence exponentially fast via the Kelvin--Helmholtz instability.
Note that this scenario produces a $-2$ spectrum in the downstream
flow, even when the shock is no longer present (resolving the first
objection mentioned in the introduction). In other words, the $-2$
spectrum is associated with the downstream flow, {\em not\/} with the
shocks themselves (as has been assumed in the past).

The second term on the right hand side of (\ref{eq:voder7}) is
baroclinic generation of vorticity due to the misalignment of pressure
and density gradients as the flow passes through the shock.  This is
the dominant term for vorticity production across straight or weakly
curved shocks.  It is also an important term in the case of multiple
shock passages as it nonlinearly mixes the flow, redistributing energy
amongst different length scales (similar the the quadratic
nonlinearity of the Navier--Stokes equations).  Note that vorticity
may be generated across a straight shock even if the upstream flow is
irrotational.

Finally, the third term on the right hand side of (\ref{eq:voder7}) is
the additional angular momentum generated by compression of the flow
in the direction normal to the front (i.e. conservation of angular
momentum). This terms simply moves the entire energy spectrum up by
the factor $\mu$ without changing its form.
 
The following sections use simple examples to show how multiple shock
passages can generate density PDFs and energy spectra similar to what
is seen in molecular clouds.  We consider three generic shock types:
weak eddy shocklets (which form spontaneously in supersonic
turbulence), focused shocks, and strong spherical shocks (which model
the blast waves generated by super novae explosions).

\section{The distribution of mass density\label{sec:density}}

In this section, we derive the density PDF of interstellar gas that
results by the passage of various shock waves.  We first demonstrate
that a log-normal distribution is very rapidly established in a medium
that is repeatedly lashed by multiple shock passages (\S~4.1).
However, in \S~4.2 we show that a power-law behaviour for a density
PDF is expected for the passage of a perfectly spherical blast wave.
Interstellar gas can therefore be regarded as being characterized by a
log-normal density PDF, which from time to time develops a power-law
tail at high densities due to the passage of a spherical blast wave
from a nearby supernova, or an ongoing stellar wind bubble.  This
situation typifies the gas dynamics in regions of star formation.

\subsection{Rapid generation of the log-normal density distribution}
We first present a very simple explanation for the origin of the
log-normal distribution of density commonly observed in isothermal
turbulent flows.  It is well-known that flows with $M_t>0.3$
spontaneously generate small, relatively weak and highly curved
shocks, called `eddy shocklets'~\cite{Kida/Orszag:1990}.  It is
therefore reasonable to assume that a region of space will be hit
several times by shocklets of varying strengths.  If we assume that
the density is approximately stationary between shock passages
(i.e. that the density changes primarily due to shock compression),
then from equation (\ref{eq:density}) the density after $n$ shock
passages is
\beq
\rho^{(n)}(x) = \prod_{j=0}^n (1+\mu^{(j)}(x))
\eeq
where we have normalized density in units of the initial uniform density
$\rho_0$.  Let us consider the shock strengths $\mu^{(j)}(x)$ to be
identically distributed random variables.  Then, since
$(1+\mu^{(j)}(x))>0$ we can take the logarithm of both sides and apply
the central limit theorem to the resulting sum.  This shows that the
logarithm of density is normally distributed, i.e. the density PDF
follows a log-normal distribution,
\beq
P(\rho) =
\frac{1}{\sqrt{2\pi}\sigma\rho}\exp\left(-\frac{(\log(\rho)-\overline{\log\rho})^2}{2\sigma^2}\right).
\label{eq:lognormal}
\eeq
Note that application of the central limit theorem to derive
(\ref{eq:lognormal}) requires only that the random variables
$\log(1+\mu^{(j)}(x))$ have finite mean and variance.

\begin{figure}
\[\includegraphics[width=0.8\textwidth,clip=true]{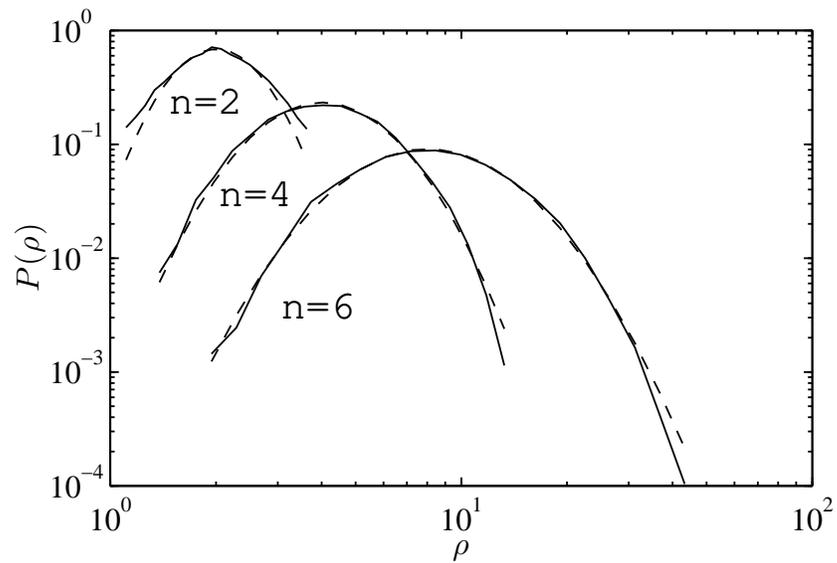}\]
\caption{Convergence to a log-normal PDF of density after $n=2,4,6$
shock passages.  - - -, log-normal distribution, --- PDF of density.
\label{fig:lognormal}}
\end{figure}

One might think that it would take hundreds of shock interactions to
converge to this log-normal distribution.  However, if the PDF of
density jumps is symmetric, then the rate of convergence is quite
fast, $O(n^{-3/2})$.  In fact, if the PDF of density jumps is uniform
(a reasonable assumption) then, as shown in
figure~\ref{fig:lognormal}, as few as three or four shock passages
generates a very good approximation to the log-normal distribution.

In particular, if the PDF of density jumps $\mu$ is uniformly
distributed in $(0,2/(\gamma-1)]$ (i.e. between its minimum and
maximum possible values), then the logarithmic mean and variance of the
log-normal distribution after $n$ shock passages are respectively 
\begin{eqnarray}
\overline{\log\rho} & = &  \frac{n}{2}\ln\frac{\gamma+1}{\gamma-1}, \\
\sigma^2 & = & \frac{n}{12}\ln\frac{\gamma+1}{\gamma-1}.
\end{eqnarray}

There are two important aspects to these results.  First, we see that
both the mean and variance of the log-normal are largest for nearly
isothermal gases (i.e. gases with $\gamma \approx 1$).  Molecular gas
can be well described as isothermal up to densities of $n \simeq 10^9$
cm$^{-3}$ due to efficient CO molecular as well as dust cooling.
Second, the result shows that the logarithmic mean increases
proportional to the number of shocks $n$ --- and that the width of the
distribution measured by the variance also grows proportional to $n$.
These trends have been observed in the density PDFs of simulations of
molecular clouds \citep[e.g., the review of][]{ MacLow/Klessen:2004}.
The overall amplitude of the distribution decreases with growing $n$
simply because the integral of $P(\rho)$ must equal unity for any $n$.
These trends are shown in figure~\ref{fig:lognormal}.  Note in
particular how few shocks are required to establish a log-normal
distribution to high accuracy.  In a self-gravitating medium such as
molecular gas, eventually gravity takes over and the dense cores begin
to collapse.

The explanation for the log-normal distribution of density proposed
here is even simpler and more general than that given previously by
\cite{Nordlund/Padoan:1999}.  In fact, our
results could explain \cite{Nordlund/Padoan:1999}'s observation that
in a numerical simulation of isothermal supersonic turbulence ``\ldots
the high-density wing of the Log-Normal is established very early --
soon after the first shock interactions.''  It is precisely these
first few shock interactions that generate the log-normal distribution.

\subsection{The power-law distribution at large densities\label{sec:dens_sph}}
We have shown that interacting weak shocklets typical of supersonic
turbulence quickly generate a log-normal PDF of mass density.
However, observations show that the PDF of mass has a power-law tail
at high masses with an exponent near the Salpeter index of
-1.35. Although the mass and density PDFs are not identical, this
observations suggests that the density PDF should also have a power
tail at high densities.  We show here that such a power-law tail may
be explained by the interaction of the log-normal density fields of
supersonic turbulence with a strong spherical shock (i.e. blast wave).

We adopt the solution for strong spherical shocks with sustained
energy injection $E(t)\propto t^p$ derived by \cite{Dokuchaev:2002}.
This solution generalizes the Sedov--Taylor self-similar solution for
a point blast explosion modelled by $p=0$ (i.e. an instant shock) to
permanent energy injection modelled by $p=1$ (i.e. an injection
shock).  The first case corresponds to the instantaneous addition of
energy to the ISM (as in a supernova explosion), while the second case
corresponds to the continuous injection of energy (as in a massive
stellar wind).  The instant shock corresponds to the classical
Sedov--Taylor solution for super novae explosions.

We assume~\footnote{We would like to thank an anonymous referee for
suggesting this space--time derivation approach.} that the PDF of
finding a particular value of gas density $\rho_1$ is proportional to
the the space--time volume where the density exceeds $\rho_1$,
\beq
P(\rho>\rho_1) \propto \int_0^{t(\rho_1)} R^3(t) \dee{t}
\eeq
where $R(t)\propto t^{(2+p)/5}$ is the radius of the spherical shock
at time $t$ and $t(\rho_1)$ is the time at which the density behind
the shock is equal to $\rho_1$.  Using the relation $M_s(t)\propto
R(t)/t$, equation (\ref{eq:density}) can be inverted to find
$t(\rho_1)\propto \rho^{5/(2(-3+p))}$.

Using the definition of the PDF, we find that the PDF of density due to
the interaction of a homogeneous gas with a spherical blast wave has
the form
\beq
P(\rho) = \frac{d}{d \rho}\int_0^{t(\rho)} R^3(t) \dee{t}
\propto \rho^{-(17+p)/(6-2p)}
\label{eq:pdf_sphere}
\eeq
(where we have re-labelled $\rho_1$ as $\rho$). Therefore, the density
PDF is a power-law with exponent $-17/6\approx -2.8$ for an instant
shock and $-9/2 = -4.5$ for an injection shock.  These slopes are
significantly steeper than the Salpeter value for the mass PDF of
$-1.35$.  However, the actual relation between the density PDF and the
mass PDF depends on the precise assumptions made about the scaling of
clumps (i.e. mass equals density times a lengthscale cubed).  Thus,
one should not necessarily expect the same index for both the density
and mass PDFs (although the power-law form should be robust).

\begin{figure}
\[\includegraphics[width=0.8\textwidth,clip=true]{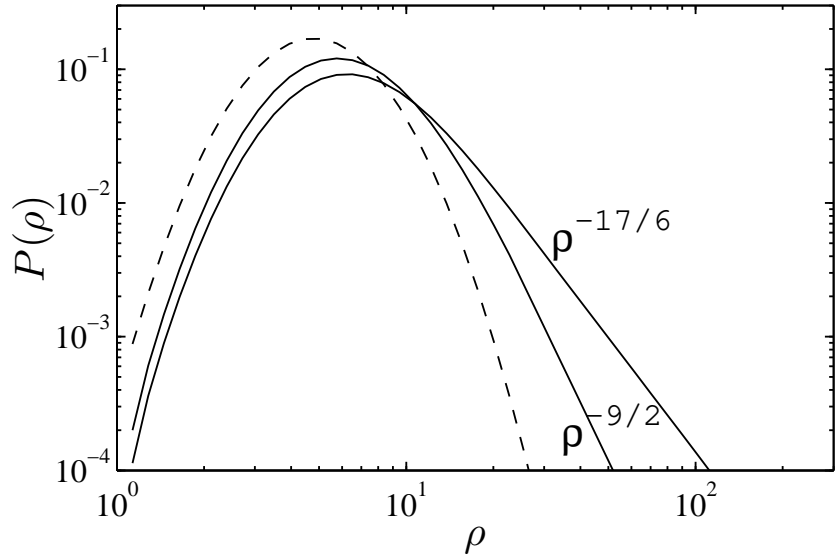}\]
\caption{Generation of a power-law PDF at large densities in a nearly
isothermal gas by a spherical blast wave interacting with a
log-normally distributed density field. - - - is the initial
log-normal PDF and the slopes -17/6 and -9/2 correspond to an instant
shock and an injection shock respectively.  Note that the upper limit
of the power-law range is proportional to $1/(\gamma-1)$, and thus we
expect the largest power-law ranges for nearly isothermal gases,
i.e. those with $\gamma\approx 1$.
\label{fig:skewed}}
\end{figure}

Mathematically, the PDF of density resulting from the interaction of a
spherical shock with a log-normally distributed density field is
simply the convolution of the PDF (\ref{eq:pdf_sphere}) with the
log-normal distribution of density (\ref{eq:lognormal}).  This
produces a PDF which is log-normal for small densities, and has a
power-law $\rho^{-(17+p)/(6-2p)}$ for high densities up to a maximum
density proportional to $1/(\gamma-1)$.  Figure~\ref{fig:skewed} shows
the resulting PDF, where the initial log-normal PDF is the result of
four shocklet passages with maximum shock Mach number $M=1.5$ in a
nearly isothermal gas with $\gamma=1.05$.  It is important to note
that because of the efficiency of CO and dust cooling, molecular gas
remains essentially isothermal to density of $\simeq 10^9$ cm$^{-3}$
--- hence our choice of an adiabatic index near to a value of unity.

Over time, the continuous production of shocklets will cause the PDF
to revert to log-normal form, at the slower rate $O(n^{-1})$ (since
the PDF is not symmetric).  Thus, the presence of a power-law in the
density PDF suggests that the flow has interacted recently with a
blast wave (such as a super nova explosion).  

\section{Energy spectrum\label{sec:spec}}

\subsection{The multi-shock model}
In addition to explaining the log-normal and power-law distributions of
density, multiple shock interactions could also explain the observed
power-law energy spectra (or velocity dispersion).  It is important to
emphasize that we consider the energy spectrum of the flow {\em
downstream\/} of the shock.  The velocity discontinuity associated
with the shock itself gives $E(k)\propto k^{-2}$, which will determine
the power-law exponent of the energy spectrum only if the downstream
flow has an energy spectrum equal to or steeper than -2 (or unless
there are no longer any shocks present).  We will show that this is
not the case in general for multiple shock passages.  To the best of
our knowledge, the special effects of focused shocks on the downstream
flow have not been examined before in the astrophysical context.

The relevance of the case of multiple shock passages has been
established by \citet{Kornreich/Scalo:2000}, who found that the average
time between shock passages in the ISM is ``small enough that the
shock pump is capable of sustaining supersonic motions against
readjustment and dissipation.''

We use a semi-analytic approach to calculate the vorticity generation
due to single and multiple passages of curved shocks.  The vorticity
jump is calculated using equation (\ref{eq:voder7}), and the velocity
and the required gradients of upstream quantities are calculated using
the fast Fourier transform (FFT) on a computational domain with
periodic boundary conditions. The computational grid is $256^3$ in all
cases.  The initial flow is assumed to be irrotational.

We make the following assumptions:
\begin{enumerate}
\item
{\bf Frozen vorticity:} flow evolution is due to the shock alone. The
flow is approximately steady between shock passages.  This is similar
to the rapid distortion approximation for strained turbulence, and it
linearizes the problem.  \citet{Kornreich/Scalo:2000} make a similar
frozen vorticity assumption to neglect flow evolution during the shock
passage. We deliberately ignore the internal dynamics of the flow:
energy re-distribution amongst scales is due to the shock.
\item
{\bf Strong shock:} the shock does not change due to interaction with the
flow. 
\item
{\bf Steady shock:} the shock's shape and strength distribution are
fixed.
\item
{\bf Random shock:} The direction and phase of the shock are chosen
randomly for each passage, and the results are averaged over many
independent realizations.  This is the same assumption we made in
deriving the density PDFs.
\end{enumerate}

This semi-analytic approach is extremely efficient numerically, and is
similar to the kinematic simulation method for
turbulence~\citep{Fung/etal:1992,Elliott/Majda:1995} In kinematic
simulation the energy spectrum is specified, but the complex phases
are chosen randomly.  Thus kinematic simulation is accurate for
quantities that depend on second-order moments of the velocity field
(e.g. particle dispersion and energy spectrum).  However, as its name
implies, kinematic simulation does not accurately represent the
dynamics of a turbulent flow.  In particular, the linearizing assumption that 
is at the heart of the model begins to break down once a significant amount
of energy piles up at smaller scales.  This will play an important role in defining
the steepness of the energy distribution, as we shall see later in this section, 
and which will be discussed in the next.

The shock profile and shock strength profile for each case are
described separately below.

\subsection{Focused shocks and shock--shocks\label{sec:shsh}}
We first consider flow driven by a {\em focused\/} shock.  As
mentioned in the introduction, a focused shock is characterized by a
flattened shock disk bounded by two shock--shocks (or a shock--shock
ring in the three-dimensional case).  The shock strength is
(approximately) discontinuous at the shock--shocks, which generate a
{\em vortex sheet\/} (with spectrum $k^{-2}$) behind the shock.
Vortex sheets are linearly unstable via the Kelvin--Helmholtz
instability, and generate turbulence very efficiently and quickly.

We model the shock profile $\phi$ and shock speed $M_s$ by
\begin{eqnarray}
\phi(x) = a \sin^4(k_1 x) \hspace*{15em}\\
M_s(x) = M_0 + 1 + \frac{4}{\pi}\sum^{n-1}_{j=1}
\frac{(-1)^{j+1}}{(2j-1)} \cos\left((2j-1) 2k_1 x\right)\mathrm{sinc}^2\left(\frac{(2j-1)2k_1}{n}\right)
\end{eqnarray}
The expression for $M_s$ is simply the Fourier series for a square
wave, where we have used the Lanczos $\sigma$-factor (the sinc term)
to remove the Gibb's oscillations.  The number of terms $N$ in the
series is taken equal to the number of Fourier modes used in the
spectral method.  We use a similar expression for a two-dimensional
shock $z=\phi(x,y)$, with wavenumbers $k_1$ and $k_2$ in the $x$ and
$y$ directions.  Although this shock is simply a model (i.e. it is not
the solution of the nonlinear wave equation governing shock motion),
it captures its main qualitative features. The shock profile and shock
strength are shown in figure~\ref{fig:focus}.

We consider two initial conditions: a uniform flow (i.e. constant density
and zero velocity), and an irrotational flow with a Gaussian energy
spectrum $E(k)\propto \exp(-k^2)$.  The latter flow models the final
decay regime of a turbulent flow, when viscous diffusion dominates.

\begin{figure}
\begin{tabular}{cc}
\includegraphics[width=0.48\textwidth]{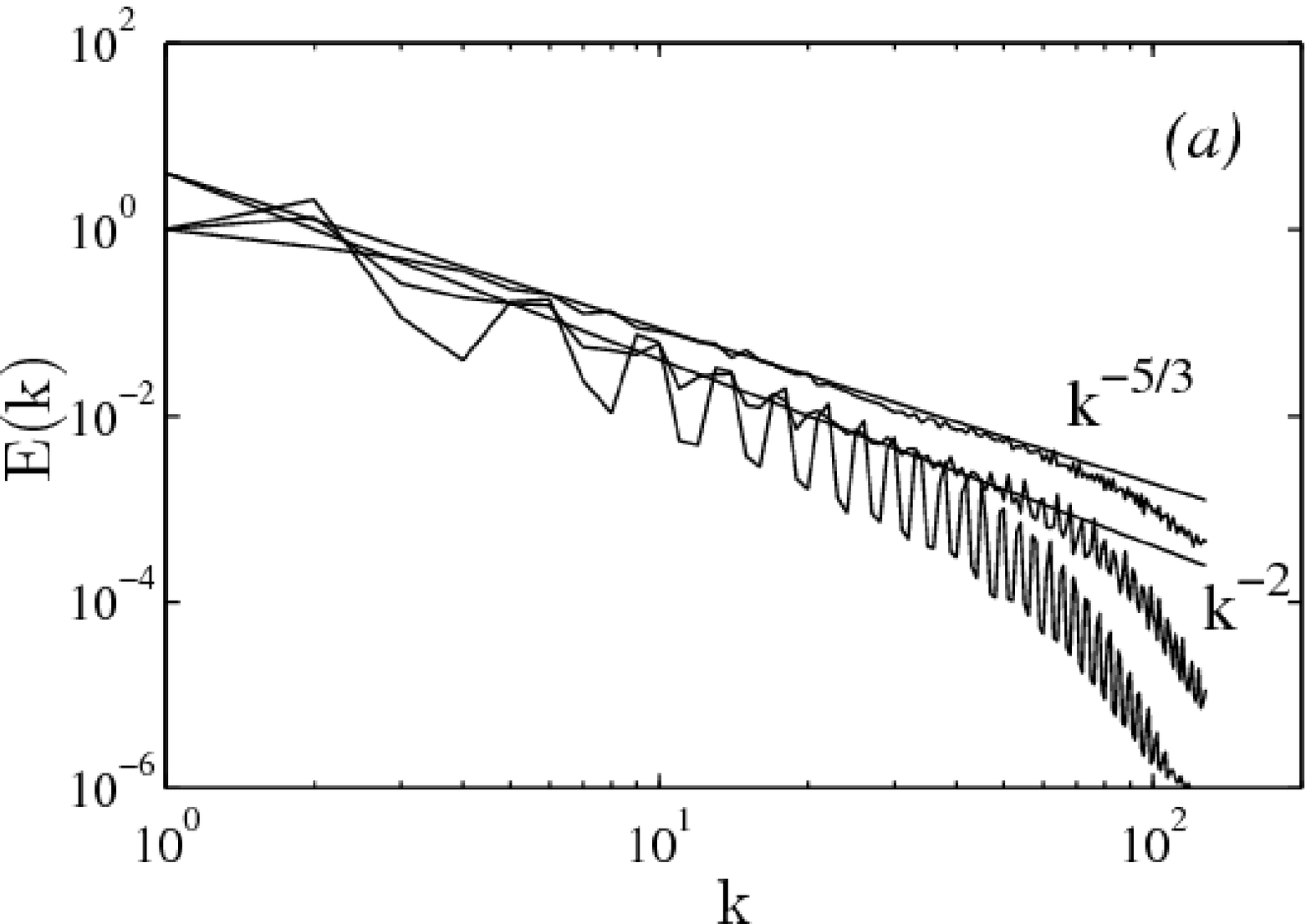} &
\includegraphics[width=0.48\textwidth]{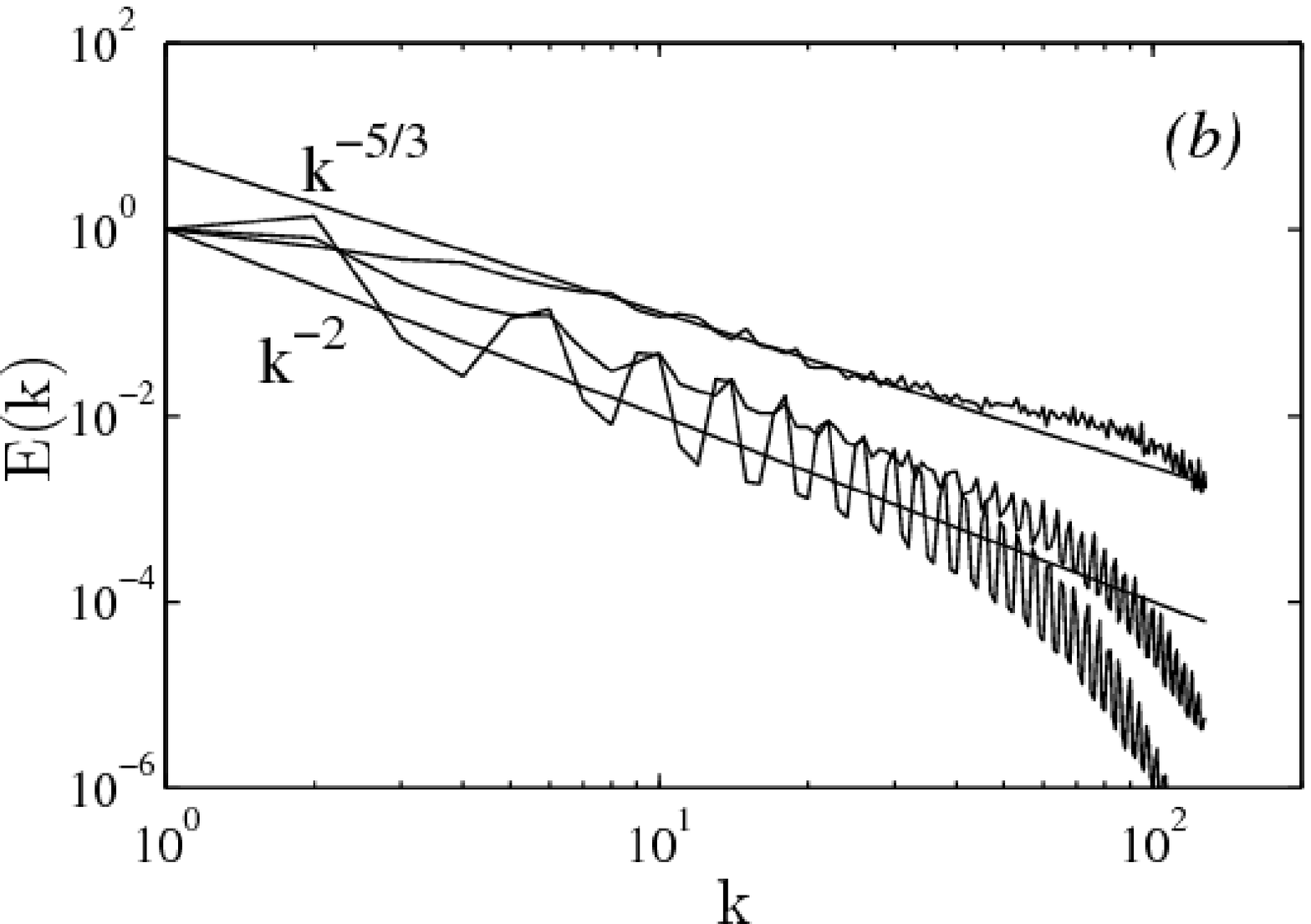} 
\end{tabular}
\caption{Energy spectra for multiple passages of focused shocks: slope
decreases with each shock passage as energy is redistributed to
smaller scales.  Then spectra have been normalized so that $E(1) = 1$.
(a)~Uniform initial flow. (b)~Gaussian initial energy
spectrum. \label{fig:focus1}}
\end{figure}

Figure~\ref{fig:focus1}(a) shows the energy spectrum of the downstream
three-dimensional flow after one, two and three passages of a focused
shock with $k_1=k_2=1$, $M_0=6$ with zero velocity initial condition.
Note that the initial $k^{-2}$ scaling of the energy spectrum (due to
the velocity discontinuity associated with the vortex sheet downstream
of the shock--shocks) becomes gradually shallower with each shock
passage.  This redistribution of energy to smaller scales is due to
the quadratic baroclinic terms depending on the (inhomogeneous)
upstream flow in the vorticity jump equation (\ref{eq:voder7}).
Although the effect is entirely kinematic, these quadratic terms
redistribute energy amongst scales in a way analogous to the quadratic
nonlinearity in the Navier--Stokes equations.

After three shock passages the energy spectrum has a scaling similar
to $k^{-5/3}$.  More shock passages would produce an even shallower
power-law.  Figure~\ref{fig:focus1}(b) shows that the form of the
energy spectrum is relatively insensitive to the choice of initial
condition.

\subsection{Spherical shocks\label{sec:sph}}
We now consider the case of perfectly spherical shocks described in
\S\ref{sec:dens_sph}.  Due to symmetry, the shock strength $M_s$ is
constant along the shock and therefore the first term in
(\ref{eq:voder7}) is identically zero.  This means that vorticity
production is due entirely to the baroclinic and angular momentum
conservation terms.  Since the initial flow is irrotational, only the
baroclinic term is active for the first shock, however subsequent
shocks generate vorticity both baroclinically and via angular momentum
conservation provided they are not coincident with the first shock.

For strong spherical shocks the vorticity jump~(\ref{eq:voder7})
reduces to
\beq
\delta\bm{\omega}\cdot\bm{b} =
\frac{2}{\gamma+1} M_s 
\left[ \frac{\partial \frac{1}{2} M_t^2}{\partial S} +
\bm{\omega}\times\bm{u}\cdot\bm{s} \right],
\label{eq:strong}
\eeq
and thus, provided the upstream flow is non-uniform, the vorticity
jump is proportional to $M_s(r)$.  

If the upstream flow is smooth and irrotational, the energy spectrum
of the downstream flow due to a single shock passage is simply the
convolution of the Fourier transforms of the singular shock strength
$M_s(r)$ given by~\cite{Dokuchaev:2002}'s blast wave solution,
\beq
M_s(r) \propto r^{-(3-p)/(2+p)},
\label{eq:Ms}
\eeq
and the gradient of the turbulent kinetic energy (e.g. $k\exp(-k^2)$).
This gives $E(k)\sim k^{-3}$ for an instant shock and $E(k)\sim
k^{-14/3}$ for an injection shock as $k\rightarrow\infty$.  The energy
spectrum for a generic self similar shock corresponding to
(\ref{eq:Ms}) is 
\beq
E(k)\sim k^{-(3+4p)/(1+p/2)}.
\eeq
Recall that these are the energy spectra of the flow {\em
downstream\/} of the shock. 

The singularity at $r=0$ is removed by using the following
regularization
\beq
M_s(r) = \frac{M_s(0)}{(1+(r/r_{min})^2)^{\alpha/2}},
\eeq
where the parameter $r_{min}$ is set slightly small than the grid size
and $M_s(0)$ is set to ensure that $M_s=1$ at the edge of the
computational domain.  The upstream flow is assumed to be irrotational
and to have a Gaussian energy spectrum $E(k)\propto \exp(-k^2)$.

\begin{figure}
\begin{tabular}{cc}
\includegraphics[width=0.48\textwidth]{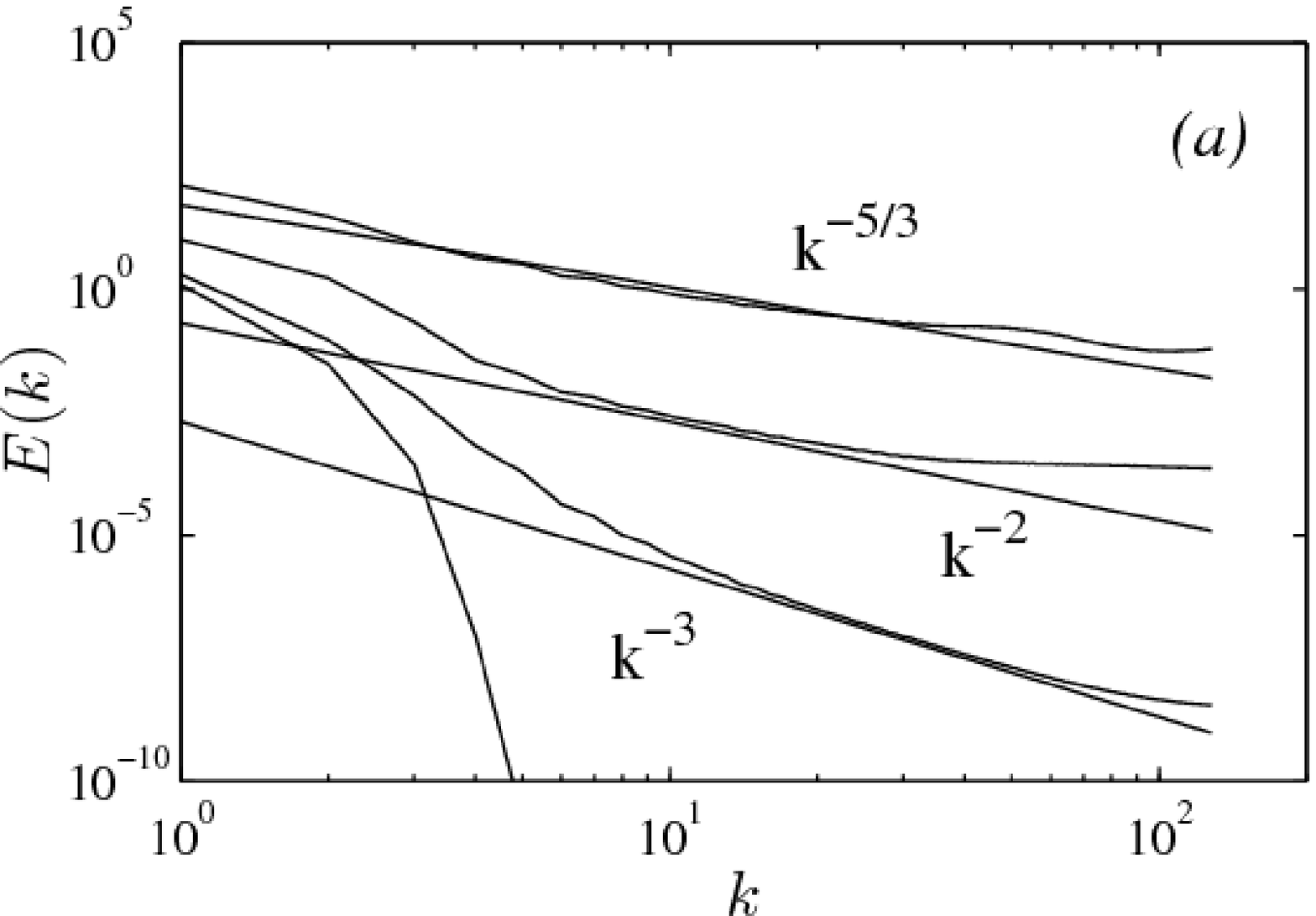} &
\includegraphics[width=0.48\textwidth]{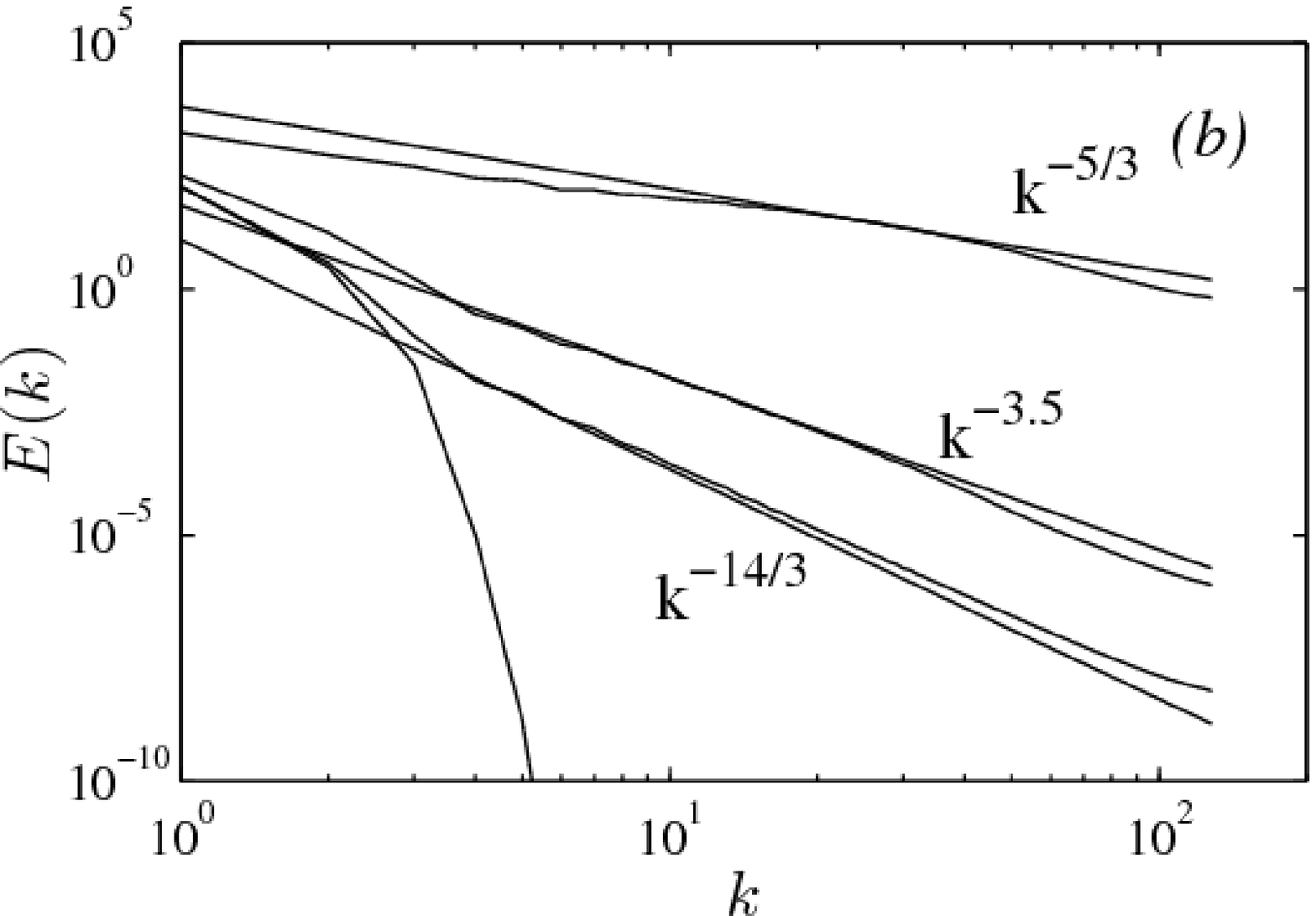} 
\end{tabular}
\caption{Energy spectra for multiple passages of spherical shocks: the slope
decreases with each shock passage as energy is redistributed to
smaller scales.  The initial energy spectrum is Gaussian and all
spectra have been normalized so that $E(1) = 1$. (a)~Instant shock
($p=0$). (b)~Injection shock ($p=1$). \label{fig:sphere1}}
\end{figure}

Figure~\ref{fig:sphere1} shows the energy spectrum of the downstream
flow after one, two and three passages of a spherical instant shock
and a spherical injection shock. Note that although the injection
shock generates a much steeper energy spectrum after one passage
($k^{-14/3}$ compared to $k^{-3}$ for the instant shock), after three
passages both spherical shock flows have a spectrum with a slope close
to $-5/3$.  As with the focused shocks, more shock passages would
further decrease the slope as energy is redistributed to smaller
scales by equation (\ref{eq:strong}).  This result is a consequence of our kinematic
treatment.  As we shall argue next, nonlinear processes will cut in to limit this 
evolution of the energy spectra.  

\section{Astrophysical implications\label{sec:concl}}
While it is well known that curved shocks are an effective way of
forcing a turbulent flow \citep{Kornreich/Scalo:2000}, we have
developed a simple kinematical model which demonstrates that shock
interaction alone may can produce energy spectra of velocity
fluctuations and mass density distributions consistent with the
observations in the ISM.  Fully developed turbulence is not necessary.
Our result has several important implications for the density
structure of gas in the ISM, and for the formation of stars within
these structures.

\subsection{Energy spectra}

Simulations have confirmed that the structure observed in the interstellar medium and molecular
clouds probably derives from shock-driven processes.  Our results show that these
processes do not drive turbulence in the classical sense of a systematic cascade in 
energy from large to small scales.
The observed $E(k)\approx k^{-5/3}$ ``big power-law in the
sky'' may actually be entirely shock-driven, and not a signature of
fully-developed turbulence.  This could explain why the power-law
extends over such a huge range of length scales (spanning several
different physical regimes, from the diffuse ISM to molecular gas),
since the power-law of a shock-driven flow is due to a singularity,
and so extends over all scales (until the viscous cut-off).  Shock driven
vortical motions are generic in the ISM, and we have shown that a log-normal ordering
of gas structure develops rapidly over all scales of the gas.  
We demonstrated that this rapid appearance of a converged log-normal
distribution is an expected property of shocked flows and the central limit theorem.

Our results also show that shocks should be much more
efficient forcing the flow than has been appreciated previously.
This is because the shock immediately distributes energy down to the
smallest scales, according to a spectrum very close to the Kolmogorov
-5/3 profile.  Relying on a classical turbulent energy cascade is not feasible, since
the time needed to transfer energy from the largest to the smallest scales
(roughly one large-scale eddy turnover time) is far too slow.

How in this picture can a turbulence spectrum close to -5/3 be universal if 
repeated shocks can continue to make it shallower?  As noted previously, 
continued steepening assumes a purely kinematic mechanism without 
the limitation of nonlinear or viscous effects.  As we pointed out in the introduction, 
although radio propagation observations find a turbulence-like spectrum close to -5/3, the full set of astrophysical
observations find spectra in the range $[-1.5, -2.6]$.  So, our
physical model needs to be able to produce a range of spectra, not just the single
universal slope.  Thus, we need to be able to suggest why -5/3 is the
most likely spectrum, as well as accounting for other slopes.

There is in fact a natural limit to how shallow the slope can be.  The
slope could never become shallower than -1, since this would imply
infinite energy (assuming an arbitrarily small minimum length scale).
With each shock passage the slope of the energy spectrum becomes
shallower as the shock redistributes energy to smaller scales (much as
the nonlinear term of the Navier--Stokes equations does).  In the
kinematic limit this process would continue until the slope approaches
-1.  However, in reality once a sufficient amount of energy
accumulates at the smallest scales energy dissipation by viscosity
becomes significant.  At this point, the linearizing assumptions of
the kinematic model break down.  Energy dissipation limits the
continued transfer of energy to smaller scales, as well as driving an
energy cascade to small scales (as in decaying turbulence).  The
energy cascade necessarily produces an energy spectrum that converges
quickly to -5/3, since we now have the conditions required for the
existence of a universal inertial range: a sink of energy at small
scales separated from a wide and continuous range of active scales.
The fact that the initial condition is already close to -5/3 means
that we expect this adjustment to happen exponentially fast (as in the
Kelvin--Helmholtz instability, where the initial spectrum is
-2). Thus, the shock forcing effectively transfers energy to smaller
and smaller scales until dissipation drives a turbulence cascade which
fixes the slope of the energy spectrum at the universal value of -5/3.

The essential point is therefore that shock-generated vorticity will
very quickly establish a power-law spectrum that is close to the
Kolomogorov value.  We argue that at this point, energy dissipation
will tend to keep it there.

\subsection{Feedback and the IMF}

Another important result of our analysis concerns the appearance of a power-law
tail for initially log-normal density PDFs that interact with spherical shock waves.  
The spherical symmetry means that a power-law will be imposed on 
the original log-normal distribution.

Consider the typical situation in a molecular cloud where a star
cluster has started to form.  We showed that the shock-driven motions
that dominate an ISM with supersonic velocities rapidly produces a
log-normal density PDF. It is well established that a typical star
forms as a member of a star cluster
\citep[e.g. reviews by][]{Pudritz:2002, Lada/Lada:2003}.  The collapse
	of the dense, gravitationally unstable regions results in the observed
IMF.  Before this process has terminated, however, a cluster is
strongly impacted by the approximately spherical shocks associated
with the most massive stars that have already formed nearby.
Observations show that most embedded clusters are adjacent to HII
regions that are excited by massive stars in nearby clusters
\citep[e.g.][]{Elmegreen:2002}.  Most clusters show evidence that
their formation could, in fact, have been triggered by the powerful
shock waves associated with the expansion of such nearby HII regions.
We showed that the passage of these shocks alters the initial
log-normal density PDF into one that has a power-law tail.  This
feedback from massive stars will therefore change the form of the IMF,
most likely by producing a power-law tail.
  
Subsequent spherical shock passages will further modify the index of
the power-law tail.  We note, however that at most one or two passages
would be expected since cluster formation is typically completed in a
million years, which is roughly the crossing time for such an event.
Thus, we conclude that feedback from massive stars is likely to leave
a signature on the form of the density PDF in the gas which will carry
over into the IMF (from those fluctuations that undergo gravitational
collapse).  Our analysis of how triggering may effect the form of the
IMF given an initial gas density PDF will appear in a future paper.

An important caveat that we have not yet discussed is the possible
role of magnetohydrodynamical (MHD) processes in shaping the density
PDFs.  Extensive sets of simulations have shown that magnetic fields
with energy densities comparable to gravitational self-energy
certainly can affect the form of the density PDF. However, the
observations are best matched with fields strengths that are much less
than this critical value. The simulations that best match the
observations of dense magnetized cores involve ``supercritical" field
strengths, i.e. fields with significantly less energy density than
gravitational and which are therefore subject to gravitational
collapse~\citep[e.g.][]{Padoan/Nordlund:1999,Tilley/Pudritz:2007,Crutcher/etal:2008}
Thus, MHD effects in the bulk of molecular gas are probably far less
pronounced than envisaged in earlier models of star formation.

Finally, and as technical aside, we suggest that it may be more useful numerically to generate a
flow with the correct energy spectrum over a very wide range of
lengthscales than to try to simulate the full nonlinear dynamics of a
turbulent flow over a very small range of length scales (as is done
currently).  For example, \cite{Elliott/Majda:1995}'s method is able
to efficiently generate a Gaussian random field with a $k^{-5/3}$
energy spectrum over 12 decades in two dimensions using only about
47\,000 computational elements (compared to $10^{24}$ computational
elements for a conventional non-adaptive approach!).  In addition to
being computationally efficient, this approach may actually be a
better model of the hydrodynamics of the ISM.

\section{Conclusions}
Our results have important implications for the origin and evolution
of density fluctuations, and particularly the CMF in molecular clouds.
Our combined log-normal and power-law distribution arises because
there are two natural kinds of symmetry to shocks ---  planar and
spherical --- that combine in a natural way.  
The fact that the power-law energy spectrum and log-normal
distribution of mass density are also observed in the diffuse ISM
strongly suggest that shock-generated vortical motions play a profound
role on all scales in the ISM, and even in star formation within 
molecular clouds.  Our specific conclusions are summarized below.  
\begin{enumerate}
\item
Supersonic turbulence with $M_t>0.3$ spontaneously generates
relatively weak and short-lived `eddy shocklets'.  A few passages of
these shocklets is sufficient to generate a log-normal distribution of
mass density.  Thus, a log-normal distribution of density should be
typical of supersonic turbulent flows, and is established very rapidly
(the passage of just a couple of shocks will suffice). This is
contrary to the usual assumption that enormous numbers of shock
passages would be required due to the slow convergence to a normal
distribution.
\item
A spherical blast wave interacting with a log-normally distributed
density field produces a power-law distribution of density at large
densities, qualitatively similar to the observed Salpeter tail for the
IMF. The power-law range increases like $1/(\gamma-1)$, and thus will
be largest for nearly isothermal gases (i.e. those for which
$\gamma\approx 1$).  Over time the power-law gradually decays to a
log-normal distribution due to the action of the continuously
generated shocklets.  Thus, the presence of a power-law in the density
distribution implies that the flow has interacted fairly recently with
a blast wave (e.g. supernova explosion).
\item
A single strong shock passage can generate a relatively steep power-law energy
spectrum over all length scales (e.g. $k^{-2}$ for a focused shock or
$k^{-3}$ for a spherical blast wave) due to the singular structure of
the shock strength.  In the kinematic limit that 
we have investigated, subsequent shock passages increase the total
energy and reduce the slope of the energy spectrum as the quadratic
nonlinear term representing baroclinic generation of vorticity by the
shock redistributes energy to smaller scales.  Three shock passages
suffice to produce an energy spectrum close to $k^{-5/3}$.  Note
that molecular clouds could not support more than a few large scale events, such 
as expanding HII regions,  without being destroyed. 

\item
We argue that the onset of energy dissipation that is expected when
energy piles up at the smaller scales acts to limit the energy spectrum
generated by shocks to the Kolomogorov value.

\item
The energy spectrum we find is that of the downstream flow, {\em
not\/} that associated with the velocity jump of the shock itself
(which has a $k^{-2}$ spectrum), i.e. we are not measuring the
spectrum of the shock itself.
\end{enumerate}
We close by noting that, to our knowledge, this is the first time
vorticity generation and mass clumping by multiple hydrodynamic, curved
shocks has been quantified analytically.

\acknowledgments
The authors wish to thank Ethan Vishniac for helpful discussions on
the observations and origins of astrophysical turbulence. We thank an
anonymous referee for very useful comments on our manuscript.  REP
also thanks Eric Feigelson for interesting discussions on the general
nature of log-normal plus power-law distributions in statistics.  The
research of both NK and REP was supported by NSERC Discovery Grants.

\bibliographystyle{apj}
\bibliography{bib_shock}
\end{document}